# Two-dimensional talc as a natural hyperbolic material

Flávio H. Feres[a#*], Francisco C. B. Maia[a#], Shu Chen[b], Rafael A. Mayer[a,c], Maximillian Obst[d], Osama Hatem[d], Lukas Wehmeier[e,g], Tobias Nörenberg[d], Matheus S. Queiroz[f], Victor Mazzotti[g], J. Michael Klopf[h], Susanne C. Kehr[d,h], Lukas M. Eng[d,h], Alisson R. Cadore[i], Rainer Hillenbrand[k,j], Raul O. Freitas[a], Ingrid D. Barcelos[a*]

This study demonstrates that two-dimensional 'talc', a naturally abundant mineral, supports hyperbolic phonon-polaritons (HPhPs) at mid-infrared wavelengths, thus offering a low-cost alternative to synthetic polaritonic materials. Using scattering scanning near-field optical microscopy (s-SNOM) and synchrotron infrared nano spectroscopy (SINS), we reveal tunable HPhP modes in talc flakes of a long lifetime. These results highlight the potential of natural 2D talc crystals to constituting an effective platform for establishing scalable optoelectronic and photonic devices.

## Introduction

Light-matter interactions at the nanometer length scale give rise to many fundamental phenomena, with polaritons[1], i.e. quasi-particles formed by the coherent coupling of light with collective oscillations, being central to these effects. Polaritons hold significant potential for applications in optoelectronics[2,3], high-resolution imaging[4,5], and chemical sensing[6,7]. Among the different types of polaritons, hyperbolic phonon-polaritons (HPhPs) stand out as a particularly intriguing class. These quasi-particles arising from the coupling of light with crystal lattice vibrations in anisotropic polar materials, are able to confine light into subwavelength volumes [8–10] and, furthermore, to spatially control and direct HPhP propagation [11–14]. Into this, polar two-dimensional materials (2DMs) demonstrate great potential for hosting such high-quality HPhPs, with their excitations typically occurring at infrared wavelengths driven by the materials' optical phononic characteristics. Notably, these phonon resonances rely on the 2DM crystalline arrangement, the type of chemical bonding, as well as the atomic mass of the constituents, which have nearly no degree of control from external stimuli[15–17].

Expanding the range of HPhPs modes across the infrared electromagnetic spectrum requires broadening the range of available polaritonic media. 2D van-der-Waals (vdW) materials, such as hexagonal boron nitride (hBN), are key examples that exhibit HPhP modes with in-plane isotropic wavefronts at mid-infrared (MIR) frequencies, located in two Reststrahlen Bands (RBs)[18]. In recent years, significant advances in exploring the polaritonic physics have been achieved by manipulating the hBN environment. These include the development of waveguided modes in hBN nanoribbons[19], controlling the group velocity of polariton pulses on engineered substrates[13,20], the nanoscale tuning of Cherenkov radiation[21], topological transitions of wavefronts in metasurfaces[22], negative polaritonic refraction[23], polariton wavelength tunability via graphene-hBN heterostructures[24–26], and the verification of strong coupling via polaritonic modes in hBN stacks[27]. Beyond the MIR, in-plane anisotropic materials such as Molybdenum trioxide (α-MoO$_3$), also show great promise for novel polaritonic applications. α-MoO$_3$ supports out-of-plane HPhP modes with elliptical wavefronts at 958 – 1010 cm$^{-1}$, and shows modes of a hyperbolic wavefront within the ranges of 545 – 965 cm$^{-1}$ [28,29]. Moreover, the HPhP response in α-MoO$_3$ has been explored also into the terahertz (THz) range of excitation[14,30]. Furthermore, in-plane anisotropic materials have also enabled the investigation of twisted optical modes, allowing to explore and quantify the topological transition of HPhP wavefronts and, consequently, to detect light channelling at MIR and THz wavelengths [31,32].

Given its wide range of applications, exploring novel platforms and materials that are capable to sustaining HPhPs across the broad excitation range from MIR to THz wavelengths. is of uttermost interests. At MIR and THz wavelengths, vanadium pentoxide (α-V$_2$O$_5$)[33] or lithium vanadium pentoxide (LiV$_2$O$_5$)[34], and germanium monosulfide (α-GeS)[35] were shown to exhibit hyperbolic HPhPs and inherent light-channelling, respectively. Additionally, hyperbolicity has been investigated in low-symmetric bulky materials, with shear and ghost polaritons being discovered in the MIR[36,37]. Semiconducting metal oxide nanowires such as tin dioxide (SnO$_2$)[12] and paratellurite (TeO$_2$)[38] were also exploited regarding their THz HPhP behaviour.

Despite these advancements, identifying a new platform for hyperbolic optics is a non-trivial task and remains challenging. Polaritonic crystals are often synthetically produced via complex chemical routes in tightly-controlled environments, with some syntheses requiring specialized instrumentation available in a few chemical laboratories worldwide. In this context, the 2D natural phyllosilicates[39] have emerged as a promising class of atomically flat, lamellar insulators[40,41]. They are vdW materials that occur abundantly in nature[42], making them readily accessible as compared to synthetically engineered polaritonic crystals. Few-layer 2D crystals obtained by conventional mechanical exfoliation offer a large electronic bandgap, as is desirable for optoelectronic and sensing applications[43–46]. Additionally, phyllosilicate crystals have a strong IR activity from far- to mid-IR[47,48] wavelengths, thus, being natural platform for phonon–polariton engineering.

a. *Brazilian Synchrotron Light Laboratory (LNLS), Brazilian Center for Research in Energy and Materials (CNPEM), Campinas, SP, Brazil.*
b. *Terahertz Technology Innovation Research Institute, National Basic Science Center—Terahertz Science and Technology Frontier, Terahertz Precision Biomedical Discipline 111 Project, Shanghai Key Lab of Modern Optical System, University of Shanghai for Science and Technology, Shanghai, China.*
c. *Institute of Physics Gleb Wataghin, State University of Campinas (UNICAMP), Campinas, SP, Brazil.*
d. *Institute of Applied Physics, Technische Universität Dresden, Dresden 01062, Germany.*
e. *National Synchrotron Light Source II, Brookhaven National Laboratory, Upton, NY 11973, USA.*
f. *Center for Research in Radiation Sciences and Technologies (CPqCTR), State University of Santa Cruz, Iheus, BA, Brazil.*
g. *Institute of Mathematics, Statistics and Cientific Computation (IMECC), State University of Campinas (UNICAMP), Campinas, SP, Brazil"*
h. *Institute of Radiation Physics, Helmholtz-Zentrum Dresden-Rossendorf, Dresden 01328, Germany.*
i. *Brazilian Nanotechnology National Laboratory (LNNano), Brazilian Center for Research in Energy and Materials (CNPEM), Campinas, SP, Brazil.*
j. *University of the Basque Country (UPV/EHU), 20018 Donostia-San Sebastian, Spain.*
k. *IKERBASQUE, Basque Foundation for Science, 48013 Bilbao, Spain .*
\# *These authors contribute equally*
\* *Corresponding authors: Ingrid.barcelos@lnls.br; Flavio.feres@lnls.br*

In this work, we demonstrate that the naturally abundant 2D crystals of 'talc', belonging to the phyllosilicate group, hold HPhP modes with circular wave fronts in the MIR range. The full optical characterization of HPhP waves in talc flakes is accomplished by scanning near-field optical microscopy (s-SNOM)[23] and synchrotron infrared nano spectroscopy (SINS)[49]. By performing real-space nanoimaging of those HPhP modes in talc that has been mounted onto various substrates such as Au and $CaF_2$, we unveil strong tunability as a function of the 2DM thickness. The extracted profiles indicate high confinement factors ($f_{con} = q_p/k_0 = 47$, where $q_p$ is the in-plane HPhP momentum, and $k_0$ is the light wavevector in the free space), a quality factor of up to 15, and a lifetime of around 2 ps. In addition, numerical simulations of the near-field distribution at the talc surface corroborate our experimental findings, adding to our comprehensive analysis. Therefore, talc constitutes a low-cost abundant natural 2D polaritonic platform in the MIR to far-IR (FIR) wavelength range.

## Experimental

### Scattering-type Scanning Near Field Optical Microscopy - (s-SNOM)

In this technique, the microscope (NeaScope, Attocube GmBH) can be described by an atomic force microscope (AFM) possessing a suited optical arrangement to acquire the optical near-field. The AFM operates in semi-contact (tapping) mode using a metallic tip that is electronically driven to oscillate (tapping amplitude of ~ 100 nm) in its fundamental mechanical frequency $\Omega$ (~ 70 kHz) near the sample surface. In our measurements, Pt/Ir coated tips with 50 nm of radius (nFTIR attocube) were used. To produce the optical near-field, the tip-sample is illuminated by an IR source that induces an optical polarization to the tip, primarily, caused by charged separation in the metallic coating, the so-called antenna effect. The optically polarized tip interacting with the sample creates a local effective polarization. The back-scattered light stemming from this tip-sample interaction is detected by a mercury-cadmium-telluride detector. The detector signal is demodulated by a lock-in amplifier in harmonics of $\Omega$. The optical near-field is given by the high harmonic (n ≥ 2) signals due to its the evanescent character. All measurements discussed here correspond to the second and third harmonics (n = 2 and 3).

In this work, for narrowband nanoimaging the s-SNOM microscope used a quantum cascade laser (QCL, model MIRCat™ from DayLight) as *quasi*-monochromatic source and pseudo-heterodyne (PSHet)[23] detection scheme. In the synchrotron infrared nanospectroscopy (SINS) experiments the tip-sample in the s-SNOM microscope is illuminated by the synchrotron broadband IR light. It is, then, performed interferometry using an asymmetric Michelson interferometer setup. The resulting interference signal is Fourier-transformed to yield the SINS amplitude $\sigma_n(\omega)$ and phase $\varphi_n(\omega)$ spectra from the complex signal $\sigma_n(\omega) e^{-i\varphi_n(\omega)}$. In the SINS data, the spectral resolution was set as 10 cm$^{-1}$ for a Fourier processing with a zero-filling factor of 4. All spectra in this work were normalized by a reference spectrum acquired on a clean gold surface (100-nm-thick Au sputtered on a silicon substrate). S-SNOM nanoimaging and SINS which were performed at the Imbuia beamline of the Brazilian Synchrotron Light Laboratory (LNLS)[50].

## Results and discussion

### Crystalline Structure and Optical Behavior

Talc, also known as soapstone[51], is an abundant, naturally occurring, layered magnesium silicate mineral[52] with the chemical formula $Mg_3Si_4O_{10}(OH)_2$. Figure 1a presents a photograph of a talc sample extracted from a soapstone mine in Ouro Preto, Brazil. Its white-green colour indicates a reduced level of impurities. Figure 1b displays layered talc flakes obtained via the standard scotch tape exfoliation method and thereafter deposited onto a non-doped silicon substrate. The exfoliation produces 2D flakes with varying lateral sizes and thicknesses, as evidenced by a flakes of various colors and contrasts, as shown in the optical image in Figure 1b. Talc has a triclinic crystal lattice formed by the intercalation of octahedral (Oc) layers of bivalent central Mg atoms with O and OH at the vertices with tetrahedral (T) layers of silicon oxide in a T - Oc- T stacking via vdW forces (Figure 1c). Its complex structure revealed a richness of vibrational modes with optical activity in the MIR[47]. The vibrational features in the MIR regime correspond to in-plane and out-of-plane Si-O stretching modes, as recently reported in References[47,48].

Due to its crystalline structure, talc exhibits isotropy in the basal plane (along the $a$ and $b$ axes) and anisotropy along the optical axis $c$) (see Figure 1c). As a result, the electrical permittivity tensor $\overleftrightarrow{\varepsilon}$ is diagonal, with distinct components: in-plane $\varepsilon_{xx} = \varepsilon_{yy}$ (where x and y correspond to the $a$ and $b$ axes, respectively) and out-of-plane $\varepsilon_{zz}$ (with the optical axis $c$ denoted as z) (Figures 1d and 1e). Using the Drude-Lorentz model (see Supporting Information for optical constants), we plot the real and imaginary parts of $\varepsilon_{zz}$ (green solid line) and $\varepsilon_{xx}$ (blue solid line), as shown in Figures 1d and 1e. The strong optical phonon activity in this range produces resonances on the $\overleftrightarrow{\varepsilon}$, leading to two well-defined Reststrahlen bands (RBs). The first RB (RB$_1$) occurs between the z-polarized transverse optical phonons ($\omega_{TOz} = 948$ cm$^{-1}$) and longitudinal optical phonons ($\omega_{LOz} = 1002$ cm$^{-1}$), as highlighted by the green shaded area in Figure 1d-e. The second RB band (RB$_2$) appears in between the in-plane polarized transverse optical phonon ($\omega_{TOx} = 1011$ cm$^{-1}$) and the longitudinal optical phonon ($\omega_{LOx} = 1041$ cm$^{-1}$). The phonon frequency values were adjusted using the Generalized Spectral Method[53] (see Supporting Information), simulating the near-field spectrum obtained from a talc crystal. The simulation accounts for possible frequency shifts in $\omega_{TO}$ and $\omega_{LO}$ due to coupling between the surface modes and the tip. This typically results in discrepancies between the far-field and near-field spectra. Within these RBs, the anisotropy in $\varepsilon$ leads to hyperboloidal-shaped isofrequency surfaces (IFCs) in momentum space ($k_z$ x $k_x$) for the electromagnetic modes in the talc crystal.

Within RB$_1$, the condition $Re[\varepsilon_{zz}] < 0$ and $Re[\varepsilon_{xx}] > 0$ results in a type I hyperbolic surface (HS), as shown in Figure 1f at $\omega = 970$ cm$^{-1}$. In contrast, in RB$_2$, the condition $Re[\varepsilon_{zz}] > 0$ and $Re[\varepsilon_{xx}] < 0$ creates a type II IFC curves, illustrated in Figure 1g for $\omega = 1030$ cm$^{-1}$. These calculations show that talc can to support high-momentum type I and II HPhP modes inside such corresponding bands. In the following, this theoretical analysis is experimentally confirmed by SINS and narrowband real-space nanoimaging of polariton waves in talc.

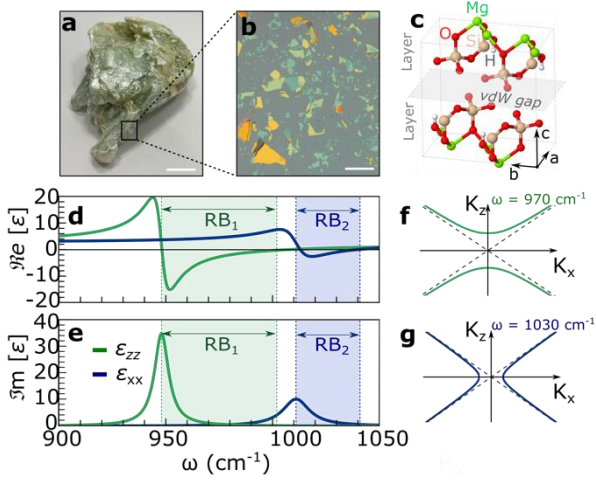

**Figure 1 | Crystalline and Optical Properties |** (a) Photography of the mineral talc block from a soapstone mine in Ouro Preto, Brazil. Scale bar of 2 cm. (b) Optical microscope images of talc flakes with varying thicknesses deposited onto a SiO$_2$(300nm)/Si substrate. Scale bar of 200 μm. (c) Schematic of the talc triclinic crystalline structure, where grey, red, green, and white spheres represent silicon (Si), oxygen (O), magnesium (Mg), and hydrogen (H) atoms, respectively. Spectral dependence of (d) the real part Re[ε] and (e) the imaginary part Im[ε] of $\varepsilon_{xx}$ and $\varepsilon_{zz}$ on talc crystal in the mid-IR range. Shaded regions indicate the two Reststrahlen bands (RB$_1$ and RB$_2$). Isofrequency curves were obtained for the frequencies 970 cm$^{-1}$ (f) and 1030 cm$^{-1}$ (g), respectively. In both frequencies, the IFC curves show a hyperbolic shape.

## Spectroscopy and Imaging of Hyperbolic Phonon Polaritons at the Mid-IR wavelengths

To investigate the HPhPs of talc mounted on top of a gold (Au), silicon (Si), and calcium fluoride (CaF$_2$) substrate, we performed both broadband spectroscopy using SINS[49] and monochromatic imaging by s-SNOM[54] (see Methods for detailed description). s-SNOM and SINS are well-established tools for probing such high in-plane momenta polaritons ($q_p \sim 10^5$ cm$^{-1}$) in the near-field zone. Importantly, the near-field tip apex supports the high-moment, since $q \sim 1/a = 3 \times 10^5$ cm$^{-1}$, which is compatible with the typical $q_p$ values in 2Ds. As a result, the s-SNOM tip can excite and probe HPhP modes via momentum conservation (Figure 2a). The excited HPhP modes, with well-defined wavelength $\lambda_p = 2\pi/q_p$, propagate through the talc sample with a circular shape due to its in-plane isotropy. The polaritonic waves are reflected by the flake edges and propagate back to the AFM tip, backscattered into the 2D sample, and finally scattered off the AFM tip into the detector. Due to the reflection at the sample edges, the HPhP modes appears with a spatial period of $\lambda_p/2$.

Firstly, using SINS we acquired a point spectrum atop of a 180 nm-thick talc flake on an Au substrate. The AFM topography of this crystal is highlighted in the inset of Figure 2b, with the green circle indicating the measurement position. The two RB bands (RB1 and RB2) are evidenced by the peaks in the normalized amplitude, $\sigma_2$. The high-amplitude peaks correspond to the optical phonons activities, which are expected for materials classified as strong oscillators in the MIR[55]. To explore the spectral and spatial behaviour within these RBs, we performed spectral line-scans using SINS. In these measurements, a set of point-spectra is acquired while the tip is scanned across the talc/Au edge (indicated by the dashed black line in Figure 2a). The linescan data are shown in Figures 2c and 2d, where the hyperbolic response of talc in both bands RB$_1$ and RB$_2$ are predicted. Moreover, polariton spatial dispersions, highlighted by the black dashed lines, are clearly visualized in both bands, albeit one can note a difference in the quality factor ($Q_f$) (defined as the ratio between the polariton propagation length (L$_x$) and wavelength ($\lambda_p$), expressed as $Q_f = L_x/\lambda_p = q_p/\gamma_p$) between the RB$_1$ and RB$_2$: type I shows three maxima, and type II only shows two maxima.

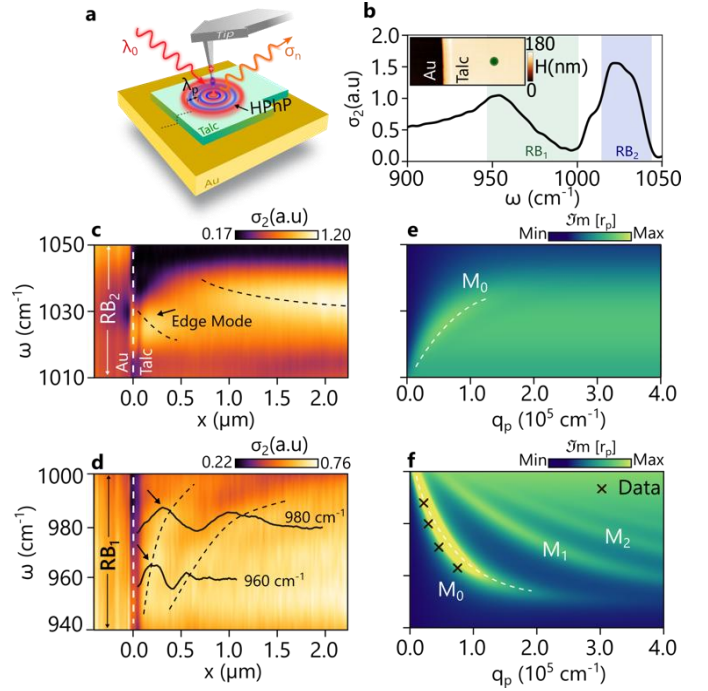

**Figure 2 | Syncrotron IR Nano-Spectroscopy of Talc crystal on Au |** (a) Schematics representation of tip-launched HPhP modes with circular wavefronts in a talc crystal on Au substrate. The AFM tip is excited by the incident light ($\lambda_0$) and the backscattered near-field electric field ($\sigma_n$) is demodulated in harmonics ($n$). The black dashed line represents the region where the spectral linescan was performed. (b) Normalized amplitude signal ($\sigma_2$) of a point spectrum (PS) acquired in a 180 nm-thick talc flake on Au using SINS. The peaks correspond to the optical phononic response in the RB$_1$ and RB$_2$ bands, highlighted by the green and blue shaded areas, respectively. The top-left inset shows the talc topography, with the pixel indicating where the point spectrum was taken. (c) and (d) Normalized amplitude signal ($\sigma_2$) of the spectral linescan performed via SINS in the talc edge, showing the RB$_2$ and RB$_1$, respectively. The vertical white dashed line marks the talc edge, while the black dashed lines represent the spatial dispersion of HPhP modes. The edge mode is highlighted by the black arrow in (c). In (d), the oscillation profiles for 980 cm$^{-1}$ and 960 cm$^{-1}$ overlay on the linescan in RB1. (e) and (f) Dispersion relation ($\omega$ - $q_p$) obtained by the plot of $\Im m[r_p]$. The white dashed lines represent the analytical dispersion performed via equation 2. The crosses (x) in (f) represent the fit-extracted $q_p$ values from the profiles.

We attribute the first maximum (closest to the edge) in RB$_2$ to edge modes, also known as Dyakonov waves[56], which are indeed expected to occur for type II modes. Excluding the first maximum from the type II HPhP profiles reveals a highly damped mode, showing one maximum in the propagation. We associate the high losses to Iron impurities in the phyllosilicate speciments[40,41,57,58], such as talc, extracted in Minas Gerais mines, in Brazil. In contrast, type I modes within the $RB_1$ show better quality factors, and no edge modes are observed. The absence of Dyakonov modes within $RB_1$ is also expected, as the out-of-plane components do not satisfy the existence condition[59]. Consequently, for type I HPhP, the first maximum can be used for polariton parameter extraction.

The real-space imaging of HPhP waves from the linescan was analysed by extracting spatial profiles (black solid lines in Figure 2d) within $RB_1$ for excitation frequencies of 960 and 980 cm$^{-1}$. As mentioned earlier, the HPhP modes exhibit an intrinsic decay, given by the convolution of two major factors: dielectric losses stemming from the material itself and the geometric spreading due to the light source in use. Tip-launched HPhP typically display geometric spreading as a function of the distance as given by $x^{-1/2}$. Thus, for extraction of the experimental frequency ($\omega$) vs. momentum ($q_p$) dispersion of talc polaritons, we model the z-component of the electric field of the HPhP modes as:

$$E_z = Ax^{-1/2}e^{i(2q_p+\varphi)x}e^{-2\gamma_p x} + B, \quad \text{eq. 1}$$

where $A$, $B$ and $\varphi$ are free parameters representing amplitude, background, and phase acquired in the reflection, respectively. Different factors, such as geometry[12] and possible heterogeneous crystalline phases (e.g., amorphous phases)[12] at the interface between air and the hyperbolic medium, can modify $\varphi$. The edge-launched polariton component can be neglected in the modelling because the tip-launching efficiency is approximately three times greater that of the edge[60]. Additionally, we consider complex momentum, $k_p = q_p + i\gamma_p$, where $\gamma_p$ represents the effective damping of the HPhP, accounting for dielectric losses and radiative scattering occurring during the reflection at the edges[61].

The fit-extracted $q_p$ (fitted profiles are shown in the SI) can be compared with the theoretical $\omega - q_p$ dispersion within the $RB_2$ and $RB_1$ (Figures 2e and 2f, respectively). As highlighted in Figure 2f, the obtained data (black crosses) show good agreement with the theoretical predictions for the type I band. The theoretical $\omega - q_p$ dispersion relations were calculated from the Fresnel coefficients for p-polarized modes ($r_p$) for a stacked system composed by air as superstrate, 180 nm-thick talc layer, and Au as substrate (see Supporting Information for more details). Figures 2e and 2f present the computed $\omega - q_p$ relations are shown as a false-color plot of the imaginary part of $r_p$ ($\Im[r_p]$)[62]. The calculated dispersion for type II HPhP modes exhibits a relatively high damping consistent with our experimental observation, where only the fundamental branch ($M_0$ - highlighted by the white dashed line in Figure 2f) is seen. It is important to note that conventional $\Im[r_p]$ formalism will not reveal the dispersion relation for edge modes[63]. However, by extracting the poles of $r_p$ and using the high-momentum approximation allows us to derive a concise analytical expression for $\omega - q_p$, as expressed by Eq. 2 (represented by the white dashed lines in Figures 2e and 2f). In that equation, $\varepsilon_\text{superstrate}$ and $\varepsilon_\text{substrate}$ are the permittivity of the superstrate and substrate layers, respectively. For our system, we define $\varepsilon_\text{superstrate} = 1$ for air, and $\varepsilon_\text{substrate}$ using the Drude model for Au (see Supporting Information). The parameter $d$ represent the Talc flake thickness, $\psi = -i\sqrt{\varepsilon_{zz}/\varepsilon_{xx}}$, while $n$ is an integer that defines the branch order. For type I bands, the $\omega - q_p$ map reveals higher-order HPhP modes ($M_1, M_2, M_3, ..., M_n$), as seen in Figure 2f. The observation of these modes in SINS and s-SNOM, however, is limited both by the lateral resolution of the tip and the material's high losses. High-order modes are typically probed in low-loss hyperbolic materials [64]. For the $M_0$ branch for type I HPhP modes, we observe a strong dependence on frequency, where the dispersion slope, i.e. the phase velocity, shows a negative behaviour[65].

$$q_p(\omega) + i\gamma_p(\omega) = -\frac{\psi}{d}\left[\tan^{-1}\left(\frac{\varepsilon_\text{superstrate}}{\varepsilon_{xx}\psi}\right) + \tan^{-1}\left(\frac{\varepsilon_\text{substrate}}{\varepsilon_{xx}\psi}\right) + n\pi\right] \quad 0, 1, 2 \ldots. \quad \text{eq. 2}$$

An estimate of the radiation confinement in talc polaritons is provided by the confinement factor $f_c = q_p/k_0 = 13$ at $\omega$ = 960 cm$^{-1}$, which is comparable to values of $f_c$ found in well-established polaritonic media as hBN and $MoO_3$[60,66]. For a broader quantitative assessment, we additionally define the quality factor, $Q_f$, a figure of merit that allows for directly comparing the performance of talc as a platform for in-plane HPhPs with other well-known polaritonic media. For Type I HPhP modes on Au, we found maximum of $Q_f = 4.5 \pm 0.5$ for 963 cm$^{-1}$ and minimum value of $Q_f = 3.6 \pm 0.3$ for 980 cm$^{-1}$. Using these extracted parameters, we calculate the polariton lifetime $\tau_s$ from $\tau_s = L_x/v_g$, where $v_g$ is the group velocity. For these modes, we found $\tau_s$ = 1.5 ps for $\omega$ = 960 cm$^{-1}$, which is comparable to the lifetime of type I HPhPs in hBN[62] and in $LiV_2O_5$[34].

In order to investigate the $\omega - q_p$ thickness dependence ($d$), we performed s-SNOM narrowband real-space imaging using the QCL source and PsHet detection (see Methods). Figures 3a and 3b display the imaging of talc on $CaF_2$ with different thicknesses (193 nm and 58 nm, respectively) illuminated at 977 cm$^{-1}$. These images reveal characteristic HPhP standing wave patterns near the crystal edges as shown by the profiles P1 and P2 extracted from Figures 3a and 3b, respectively. For a more complete evaluation of talc HPhPs as a function of thickness, we measured different crystals by varying the illuminating frequencies and plotted the dispersions (Figure 3e and 3f).

The fit-extracted $q_p$ values for different frequencies and flake thicknesses are plotted in Figure 3e, matching the predicted behaviour from the analytical dispersion relation (Equation 2), where $q_p \propto 1/d$. For thinner flakes, we observed higher $f_c$ values, reaching a value of 47 at 966 cm$^{-1}$ for a thickness of 58 nm. In general, within the examined momentum range, the tip-probed polariton modes emerge near the efficiency peak at $q \sim 1/a = 3 \times 10^5$ cm$^{-1}$. However, for thinner flakes ($d < 50$) where $q_p$ increases, it becomes increasingly challenging to probe HPhP modes using s-SNOM due to the momentum mismatch between the tip and the HPhPs. Based on this analysis, we identify the range of $d \sim 50 - 200$ nm as optimum for HPhPs interferometry by s-SNOM in talc flakes, independently of the substrate. The substrate permittivity ($\varepsilon_\text{substrate}$) also plays a significant role in the tunability of $q_p$. For example, the $q_p$ values for a 180 nm-thick flake on an Au substrate (where $\Re e[\varepsilon_\text{Au}] = -5000$) are shorter than those observed for a 193 nm-thick flake on a $CaF_2$ substrate (where $\Re e[\varepsilon_{CaF_2}] = 1.14$). Typically, type II HPhP modes present larger wavelengths on dielectric substrates than metallic ones, while the opposite is observed for type I HPhP modes[13,67]. This behaviour, which was used to accelerate polaritons via modification of the dielectric environment, has been documented in the literature for various systems[13].

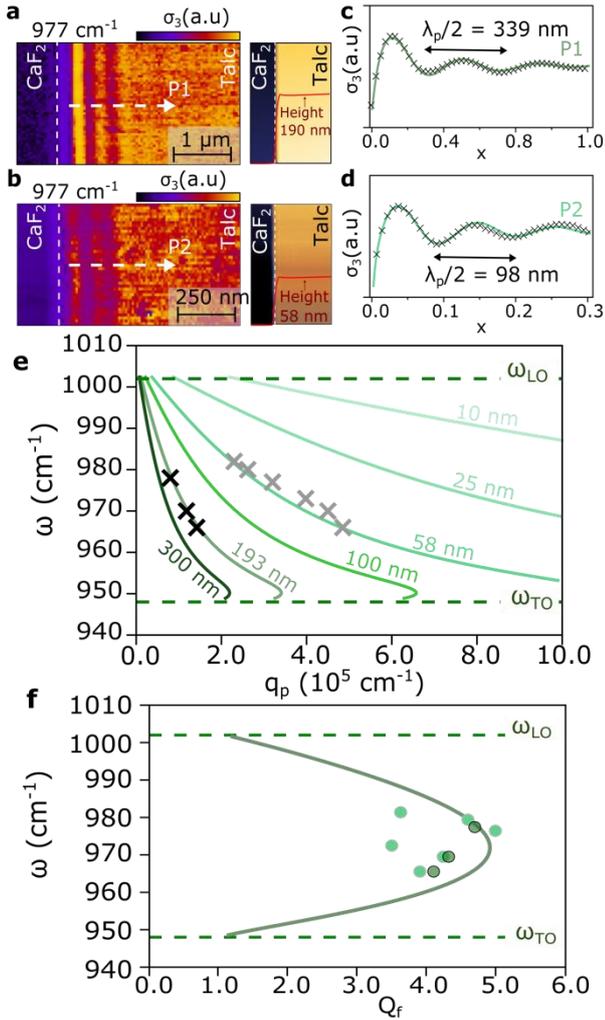
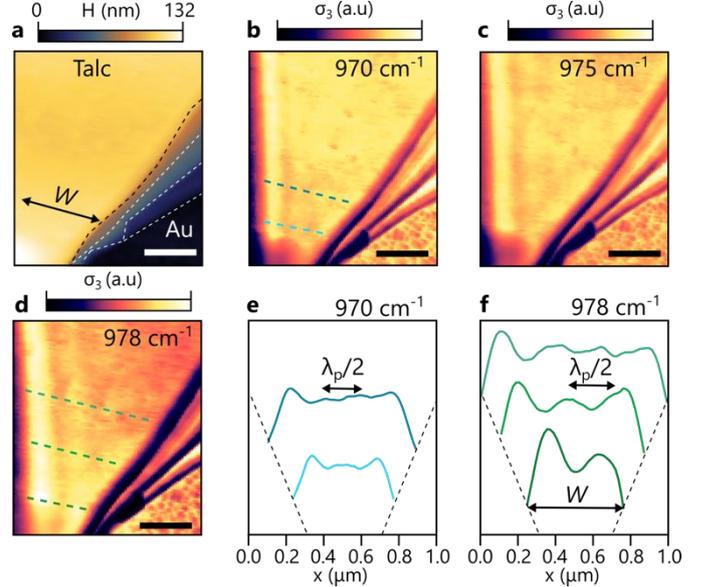

the flake forms a triangular cavity with a variable width ($W$), where the FP cavity polaritonic mode arise from multiple reflections as presented in Figures 4b-d from imaging at 970 cm$^{-1}$, 975 cm$^{-1}$, and 978 cm$^{-1}$. The standing wave pattern at 970 cm$^{-1}$ exhibit a varying number of maxima as the cavity width $W$ increases. Additionally, changes in the standing wave pattern at other frequencies are observed due to the dependence of the wavevector $q_p$ on the frequency $\omega$.

**Figure 3| Real-space narrowband nano-imaging of talc flakes on CaF$_2$ substrates|** (a) and (b) s-SNOM nano-imaging and AFM topography of 193 nm and 58 nm-thick talc flake on CaF$_2$ substrate. The height profiles are represented by the red solid line. The optical images display the $\sigma_3$ acquired with an excitation of 977 cm$^{-1}$. The profiles P1 and P2, extracted from (a) and b), are shown in (c) and (d), respectively. (e) Analytical dispersion relations ($\omega - q_p$) obtained for various talc thicknesses (300 nm, 190 nm, 100 nm, 58 nm, and 25 nm). The symbols (x) represent the experimental $q_p$ values obtained from fitting. (f) Quality factor ($Q_f$) curve for the RB$_1$ HPhP modes. The green solid line shows the theoretical prediction, while the symbols represent experimentally extracted values.

The experimental $Q_f$ values obtained for talc on CaF$_2$ are consistent with those on Au, aligning well with theoretical predictions (solid green line in Figure 3f). Thus, we note that $Q_f$ is substrate-independent. With $Q_f$ values fluctuating between ~3 and 5, our findings suggest that talc is a promising, low-cost, and naturally abundant platform for polaritonic applications.

Given the observed $Q_f$'s values, one promising application is the creation of polaritonic Fabry-Perot (FP) micro- and nanocavities of talc crystals. An example is demonstrated in Figure 4, where a set of narrowband s-SNOM images of a triangular-shaped 163 nm-thick talc on Au substrate unveil HPhP cavity modes. Figure 4a shows the AFM topography, with dashed lines indicate the crystal region of constant thickness where the modes rise. As commented, the flat region of

**Figure 4| Imaging of cavity modes in talc|** (a) AFM topography of a 132 nm-thick talc crystal on Au substrate. The dashed lines mark the crystal boundaries, indicating regions of varying thickness. The top portion of the flake forms a triangular-shaped cavity with variable width $W$. (b) – (d) Amplitude signal ($\sigma_3$) acquired via s-SNOM at the frequencies 970 cm$^{-1}$, 975 cm$^{-1}$, and 978 cm$^{-1}$, respectively. The scale bars represent 400 nm. The blue and green dashed lines in (b) and (d) indicate the locations where the profiles were extracted, as shown in (e) and (f). The standing wave pattern changes with varying W, and the distance between consecutive maxima corresponds to the spatial period $\lambda_p/2$.

In Figures 4e and 4f, we exhibit the profiles marked by the dashed lines in the images at 970 cm$^{-1}$ and 978 cm$^{-1}$, respectively. The profiles reveal a highly symmetric pattern between the boundaries, indicating that the truncated geometry facilitates constructive interference, which satisfies the FP condition:

$$q_p W + \phi = (n+1)\pi \qquad \text{eq. 3}$$

where $\phi$ represents the phase accumulated during the round-trip multiple reflections, and $n$ is the mode order. The resulting mode exhibits a spatial period of $\lambda_p/2$, from which the $q_p$ values can be determined by measuring the distance between two consecutive maxima.

## Conclusions

We introduce two-dimensional talc crystals as a novel and promising platform for hyperbolic phonon polariton (HPhPs) generation at mid-infrared (MIR) wavelengths. Using both scattering-type scanning near-field optical microscopy (s-SNOM) and synchrotron infrared nano-spectroscopy (SINS), we confirm the presence of both Type I and Type II HPhP modes with circular wavefronts in talc crystals. Notably, Type I modes exhibit lower losses as compared to type II modes, with significant lifetimes ($\tau_s \sim 2$ ps) and high confinement factors ($f_c \sim 13$), comparable to well-established polaritonic materials. Our experimental data align closely with theoretical predictions, revealing the tunability of the HPhP behaviour across different substrates and talc sample thicknesses. One of the outstanding features of talc is, unlike synthetically engineered polaritonic crystals, its natural abundance and low cost, thus offering a practical and sustainable platform for large-scale applications. This affordability and ease of acquisition position talc as a key contender in the phonon-polaritons field without complex fabrication processes.

Moreover, we also identify an optimal talc thickness range for in-plane polaritonic applications of $d \sim 50 - 200$ nm, applicable when using both metallic and dielectric substrates. The HPhP modes achieve a maximum quality factor of $Q_f \sim 5$, allowing for the direct observation of standing wave patterns near the crystal edges, which enables the formation of Fabry-Perot cavity modes for selected geometries. In particular, we observe Fabry-Perot modes along the edges of triangular talc crystals, similar to those reported for graphene monolayers and hBN layered crystals. Additionally, we also found evidences of HPhP modes in the far-infrared (FIR) range of excitation when using the FEL/s-SNOM setup (see Supporting Information), although with lower $Q_f$ values. These findings position talc as an ultra-broadband platform for polaritonic applications, heavily expanding its capabilities and use as compared, for instance, to α-MoO₃ crystals. The combination of talc's naturally abundant, low-cost nature with its tuneable polaritonic properties opens up new avenues for both fundamental research and practical, scalable applications in MIR-to-THz photonics and optoelectronics.

## Acknowledgements

All Brazilians also thank the Brazilian Synchrotron Light Source (LNLS) for providing beamtime for the experiments and the Brazilian Nanotechnology National Laboratory (LNNano), both part of the Brazilian Centre for Research in Energy and Materials (CNPEM), a private non-profit organization under the supervision of the Brazilian Ministry for Science, Technology, and Innovations (MCTI), for sample preparation and characterization – IMBUIA (Proposals 20232578), LNNano (Proposals: 20240216) and LAM (Proposals: 20240702) at LNLS. Parts of this research were carried out at the ELBE Center for High-Power Radiation Sources at the Helmholtz–Zentrum Dresden–Rossendorf e. V., a member of the Helmholtz Association. F.H.F. and R.A.M acknowledges FAPESP support through the grant 2023/09839-5 and 2020/15740-3, respectively. I.D.B. acknowledges the prize L'ORÉAL-UNESCO-ABC for Women in Science Prize – Brazil (2021 edition). F.C.B.M., I.D.B. and A.R.C. acknowledge the FAPESP financial support (2022/02901-4). I.D.B., R.O.F., A.R.C and F.C.B.M acknowledge the CNPq through the research grant 306170/2023-0, 309946/2021-2, 309920/2021-3 and 313672/2021-0, respectively. R.O.F. acknowledges the support from FAPESP Young Investigator grant 2019/14017-9. I.D.B. and A.R.C. acknowledge the financial support from the Brazilian Nanocarbon Institute of Science and Technology (INCT/Nanocarbono). F.H.F., L.W., M.O., O.H., T.N., J.W., F.G.K., S.C.K., and L.M.E. acknowledge the financial support by the Bundesministerium für Bildung und Forschung (BMBF, Federal Ministry of Education and Research, Germany, Project Grant Nos. 05K16ODA, 05K19ODB, and 05K22ODA) and by the Deutsche Forschungsgemeinschaft (DFG, German Research Foundation) through the project CRC1415 as well as under Germany's Excellence Strategy through Würzburg-Dresden Cluster of Excellence on Complexity and Topology in Quantum Matter—ct.qmat (EXC 2147, project-id 390858490). L.W. acknowledges funding by the U.S. Department of Energy, Office of Science, National Quantum Information Science Research Centers, Co-design Center for Quantum Advantage (C2QA) under contract number DE-SC0012704."